# The Improvement of Joule Balance NIM-1 and the Design of New Joule Balance NIM-2

Z. Li, Z. Zhang, Q. He, B. Han, Y. Lu, J. Xu,
S. Li, C. Li, G. Wang, T. Zeng, and Y. Bai

**Abstract**—The development of the joule balance method to measure the Planck constant, in support of the redefinition of the kilogram, has been going on at the National Institute of Metrology of China (NIM) since 2007. The first prototype has been built to check the feasibility of the principle. In 2011, the relative uncertainty of the Planck constant measurement at NIM is $7.7 \times 10^{-5}$. Self-heating of the coils, swing of the coil, are the main uncertainty contributions. Since 2012, some improvements have been made to reduce these uncertainties. The relative uncertainty of the joule balance is reduced to $7.2 \times 10^{-6}$ at present. The Planck constant measured with the joule balance is $h=6.6261041(470) \times 10^{-34}$Js. The relative difference between the determined $h$ and the CODATA2010 recommendation value is $5 \times 10^{-6}$. Further improvements are still being carried out on the NIM-1 apparatus. At the same time, the design and construction of a brand new and compact joule balance NIM-2 are also in progress and presented here.

## I. INTRODUCTION

THE kilogram is the last one of the seven SI base units that is still kept by an artifact, the International Prototype of Kilogram (IPK). Comparisons showed that there is an obvious drifting among IPK and its official copies [1] as shown in Fig 1.

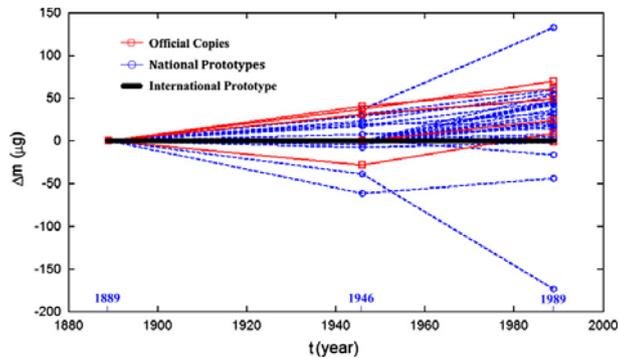

Fig.1 The comparison between IPK and 6 official copies

In the comparison, the IPK is assumed to be stable. The difference between IPK and the official copies are measured and a drift of 50μg in 100 years is

Manuscript received Aug. 26, 2014. This work was supported in part by the National Natural Science Foundation of China (51077120), and the National Department Public Benefit Research Foundation of China (201010010).
Z. Li is with the National Institute of Metrology (NIM), Beijing 100029, China (phone: +86-10-64526175; e-mail: lzk@nim.ac.cn).Z. Zhang, Q. He, B. Han and Y. Lu are with the NIM. J. Xu, S. Li, and C. Li are with Tsinghua University, Beijing 100084, China. G. Wang is with Beihang University, Beijing 100191, China. T. Zeng and Y. Bai are with Harbin Institute of Technology, Harbin, 150001, China observed. Since the same material is used for the IPK and official copies and the six official copies are also kept in the same protective conditions as IPK, it is not reasonable for the assumption that the IPK never changed in the last 100 years. Being lack of an absolutely stable reference, the actual drift rate cannot be known. A quantum mass standard is anticipated to be established to monitoring the actual drift of the IPK and further replace the artifact standard of mass [2]. Since 1970s, great efforts have been put forward to the research on several different approaches, such as voltage balance, watt balance, Avogadro project [3]. To avoid possible system errors from one method, more experiments especially with different principles are expected and encouraged for the final determination of the Planck constant. To make contribution for the redefinition of kilogram, the National Institute of Metrology of China (NIM) proposed a joule balance method in 2006 [4], which is also an electrical way but different from watt balance method [5]. The principle of the joule balance is shown in Fig. 2 and briefly introduced late [6, 7].

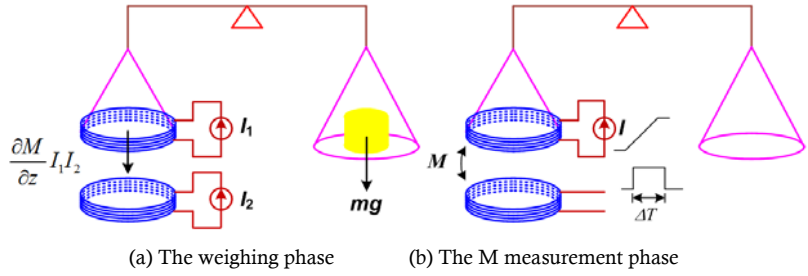

(a) The weighing phase  (b) The M measurement phase
Fig.2 The principle of the joule balance

There are two phases in a whole measurement procedure, i.e. weighing phase and mutual inductance measurement phase. In the weighting phase as shown in Fig. 2(a), currents are applied to two coil sets, one of which is fixed to the base of the balance and the other is suspended to one arm of the balance. Then we can have,

$$mg = \frac{\partial M}{\partial z} I_1 I_2 + \Delta f_z \quad (1)$$

Here,
 - $m$ is the standard mass;
 - $g$ is the gravitational acceleration;
 - $M$ is the mutual inductance between two coil sets;
 - $z$ is the position of the suspended coil in the vertical direction;
 - $I_1$ and $I_2$ are currents applied to two coil sets;
 - $\Delta f_z$ is a very small residual force in weighing phase.

To avoid the difficulty of determining the derivative quantity $\partial M / \partial z$ directly, by integrating both sides of equation (1) along the vertical track from $z_1$ to $z_2$, we can get,

$$mg(z_2 - z_1) = [M(z_2) - M(z_1)] I_1 I_2 + \int_{z_1}^{z_2} \Delta f_z(z) dz \quad (2)$$

Thus only two quantities, mutual inductance values $M(z_1)$, $M(z_2)$, and a relative distance $(z_2-z_1)$, should be measured at static situation to determine $m$. The dynamic measurement procedure in the watt balance method is avoided. It can be seen that (2) is an equation between gravity potential and electromagnetic energy, which is the origin of 'joule balance'.

In traditional method, the uncertainty for determining the mutual inductance is at $10^{-6}$ level. Thus, it is obvious that the mutual inductance is one of the main difficulties limiting the joule balance method. With years of efforts, a standard square wave compensation method has been proposed by authors [8] and the uncertainty is at $10^{-7}$ level at present. When PJVS is used for the mutual inductance measurement system, the uncertainty can be improved further.

The joule balance project got funds and was started in the early of 2007. The first prototype was build at the end of 2011 to check the principle of this new approach. The relative uncertainty of the Planck constant measurement result is several parts in $10^5$ as shown in Table I.

TABLE I
THE UNCERTAINTY BUDGET OF THE JOULE BALANCE NIM-1

| Parameter | Type | Uncertainty ($k$=1, $10^{-6}$) |
|---|---|---|
| Mutual inductance measurement | A, B | 0.1 |
| Current sensing | B | 2 |
| Self heating of the coil | A, B | 56 |
| Force weighing | A, B | 20 |
| Mass measurement | B | 5 |
| Length measurement | A, B | 49 |
| Gravity | A, B | 0.01 |
| Air buoyancy | A, B | 1 |
| **Combined uncertainty** | **A, B** | **77** |

The uncertainty is mainly from the following aspects:

*A. Self-heating of the Coils*

The resistances of the fixed exciting coil and the suspended coil are 1.6k ohm and 130 ohm respectively. In the weighing mode, 250mA generated by two current sources flow through both coils, thus producing power of 100W and 8W respectively, which introduce several big problems, such as expansion of coils' sizes, air fluctuation from the temperature gradient and heating the position sensors of balance above the fixed coils.

*B. The Reading of the Balance*

To decrease the horizontal movement of the suspended coil when it is moved to different position in vertical direction, a balance with a 2-meter long beam is used. However, the long beam would be susceptible to the disturbance of either airflow or vibration from surroundings. Besides, the position sensor of the balance is just above the coils and the heating from the coils will cause the position sensor to drift. The resolution of the balance is 1 mg and repeatability is 10mg and it introduces 20ppm uncertainty contribution.

*C. The Swing of the Suspended Coil*

The self oscillation of the beam and the air fluctuation cause the suspended coil swing. The laser system for the length measurement cannot work well in this situation. Since 2012, several improvements have been done to decrease these uncertainties. Details of the improvements are discussed in section II. Due to the big size of NIM-1, vacuum is impossible and the length measurement is limited at several parts in $10^6$. Thus, the combined uncertainty of the Planck constant is limited to $10^{-6}$ level. To get a result at several parts in $10^8$ and fulfill the requirement of Consultative Committee for Mass and Related Quantities (CCM) [9], a brand new, compact and vacuum compatible joule balance NIM-2 is in design and construction. Details are presented in section III.

## II. THE IMPROVEMENTS OF THE JOULE BALANCE NIM-1

To decrease the large uncertainty contributions in table I, since 2012, some improvements have been done. At the same time, these approaches for the improvements are also helpful for the design of the new joule balance NIM-2. Details of the improvements are described as following.

*A. Decreasing the Self-heating of the Coils*

In the first version of the joule balance apparatus finished by the end of 2011, two coil sets wound with enameled copper wire are used to generate the electrical force to balance the gravitational force of the mass [6]. However, the self heating of the coils with currents brings some trouble. The power for two coil sets is more than 100W. Experiments show that the heating will increase the airflow and introduce noise and drifting for the readings of the balance. Later, a superconducting coil has been constructed to avoid the heating problem.

In the mutual inductance measurement phase, for the coils wound with enameled copper wire, the primary of the mutual inductor is the suspended coil with just one coil, and the secondary is the exciting coil which consists of two coils connected in series-opposition, thus reducing the coupling from outer magnetic field. But it is different when the superconducting coil is used. To avoid eddy current effect, the Dewar for the superconducting coil is made of fiberglass reinforced plastic (FRB). A cold head is mounted on the top of Dewar and some corrugated pipes from the cryocooler are also connected with it. It is very heavy and is not easy to drive it up and down. Thus, the superconducting coil is used as the primary of the mutual inductor and the exciting coil, and the secondary is the suspended coil set with two coils wound with enameled copper wire connected in series-opposition.

But experiments show that the Meissner effect of superconducting coils will cause a troublesome measurement uncertainty [10]. The magnetic status is different for the weighing phase and the mutual inductance measurement phase. In the first phase the movable coil hang on the balance beam is with a current to produce the magnetic force, but in the second phase no current is applied for this coil, and only the induced voltage of the suspended coil is measured. Thus the current distribution in the fixed superconducting coil will be changed due to Meissner effect during the measurement procedure turns from one phase to another. Some systematic error will take place and it is difficult to evaluate them precisely.

Then we had to go back to the coils system wound with enameled copper wire again. To decrease the big uncertainty from the self heating of the coils with currents, an optimized design of the coils system was proposed to decrease the turns of the coil but keep the electrical force almost the same. Fig.3 shows the optimized coils.

The distance from exciting coil to the movable coil is decreased. The calculation and test results show that this compact design reduce the coil heating power to 1/3 compared with the formerly used coil sets [11].

By measuring the electrical force generated with the compact coil sets at different currents and linear extrapolation, the uncertainty of the self heating of the coil can be evaluated. The uncertainty contribution is $7\times10^{-6}$ at present.

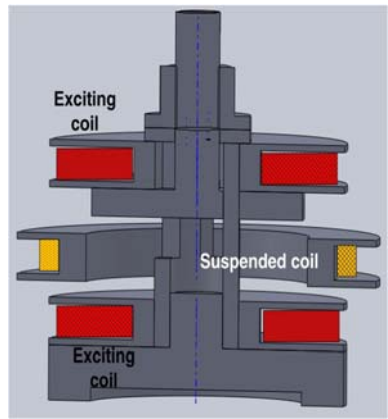

Fig. 3. The compact coil set

*B. A linear driving stage is used to move the exciting coil*

At the beginning, by tilting the balance arm to different angle, the suspended coil is moved to different positions in vertical direction to perform the mutual inductance measurement. But the horizontal movement of the suspension introduces the swing of the suspended coil. A new approach is used since 2013, i.e., the suspended coil is kept at a stable position and the exciting coil is moved up and down. A linear driving stage is developed to move the exciting coil up and down in a 20mm range in the vertical direction. The relative difference of suspended coil and exciting coil is measured. Measurement results show that the linearity of the driving stage is about 10um and will be improved further.

*C. The Laser Based Position Locking System of the Suspended Coil*

The self heating of the coils will increase the temperature gradient and introduce the air convection. The air convection will further bring some force in both of the vertical and horizontal directions and contribute to the swing and vibration of the suspended coil. When the mass is put on and moved away the suspension, this action will also introduce some disturbance to the stability of the suspended coil. Three oil damping devices are used to decrease the swing. The long beam of the balance also introduces some mechanical oscillation.

A laser heterodyne interferometer is incorporated into a three-axis simultaneous differential interferometer to measure the relative displacement between the suspended coil and the exciting coil [12] as shown in Fig.4 . Measurement results show that vibration and conical swing are observed. At the same time, an obvious drifting, which is mainly from the self heating of the coils, is observed. Fig. 5 shows the vibration and drifting of the suspended coil.

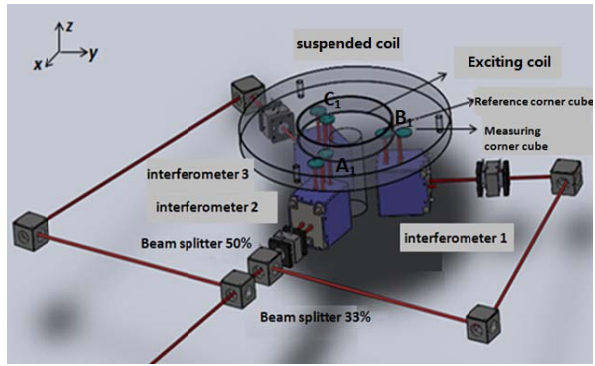

Fig. 4 The laser system for position locking of the suspended coil

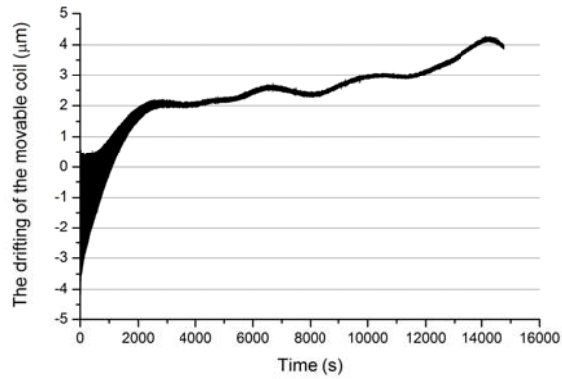

Fig. 5. The drifting of the suspended coil

To keep the suspended coil stable, three oil damping devices are used to decrease the swing along the balance beam and the conical swing [12]. An active control technique to decrease the vibration and drift of the suspended coil is also proposed by authors and has been verified with a model [12]. The principle of the system is shown in Fig.6.

The key point is the using of a piezoelectric ceramic control unit to compensate the drift and low-frequency vibration of the suspended coil, thus keep it at a stable position. The piezoelectric ceramic is controlled by a PID controller, which collect the moving information of the suspended coil with a laser interferometer system. This approach has been used in NIM-1. Fig. 8 shows that it reduces the random coil position variation and drifting greatly.

In a whole measurement procedure, the total distance of the exciting coil is 20mm. The relative distance change between the suspended coil and the exciting coils is measured. With the laser based position locking of suspended coil, the accuracy of the length determination can be decrease to 10 nm level. The uncertainty for length measurement is improved to $0.5\times10^{-6}$.

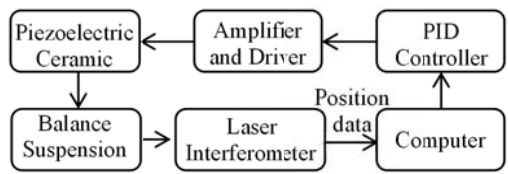

Fig.6 The block diagram of the control principle

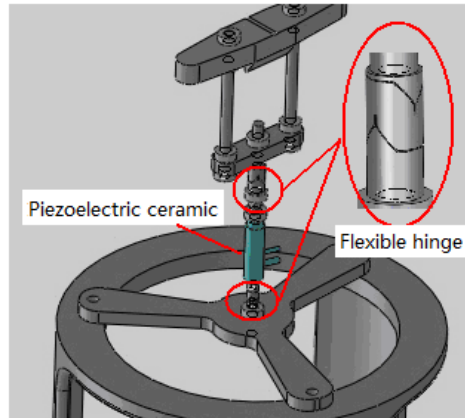

Fig.7. The piezoelectric ceramic used in the laser based position locking system for the suspended coil

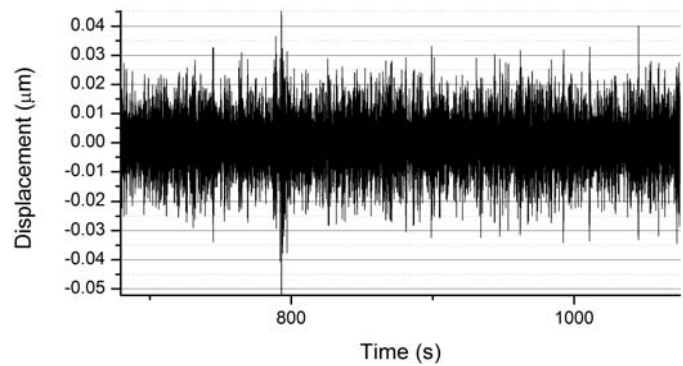

Fig. 8. The stability of the suspended coil with the laser based position locking system and oil damping devices

*D. Mass Measurement*

An E2 class mass of 200g traced to the primary standard of mass is used to replace the formerly used copper mass. The uncertainty from the calibration report is 0.1mg ($k$=2). Thus the relative uncertainty of mass measurement is $0.25 \times 10^{-6}$ ($k$=1).

*E. Improvement of the Control System of the Balance*

For the control system of the original version of the balance system, a voltage source is used. When current flows in the actuator coil, the resistance will increase and the reading of the balance will drift too. The noise in the coil will also introduce some noise in the reading of the balance.

A current source is developed to replace the voltage source and the drifting of the reading is decreased a lot.

With this approach and oil damping device mentioned above, the repeatability of force weighing has been improved to about 2mg, the relative standard uncertainty for the force weighing is reduced to $1 \times 10^{-6}$.

*F. Alignments*

To estimate the misalignment error caused by the horizontal movement of fixed coil moved by the linear driving stage mentioned above, a high precision measuring method based on self-alignment capacitive displacement sensor is proposed to measure the displacement of the joule Balance in the horizontal direction. A thin copper wire with a plumb is used as the alignment reference of the gravity direction as shown in Fig.9. The capacitor is formed by a thin copper wire and a copper cylinder. The capacitance between them is monitored to get the horizontal displacement of the exciting coil when it is moved up and down. Experiment show that the resolution of the senor is 1 μm and is being improved further [14]. Besides, to tune the alignment of the linear driving stage, a 3-dimensions adjustable base is installed.

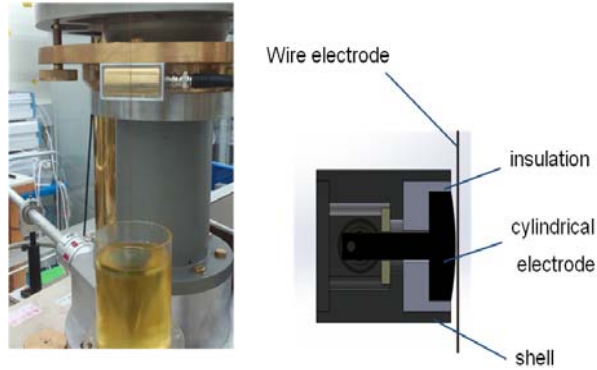

Fig.9. The alignment approach used in NIM-1

### G. Uncertainty Budget

After these improvements, the uncertainty of the joule balance is at $7.2\times10^{-6}$. The Planck constant measured with joule balance is $h=6.6261041(470)\times10^{-34}$Js. The relative difference between it and CODATA2010 value is $5\times10^{-6}$.

TABLE II
THE UNCERTAINTY BUDGET OF NIM-1 AFTER IMPROVEMENTS

| Parameters | Type | Uncertainty ($k$=1, $10^{-6}$) |
|---|---|---|
| Mutual inductance measurement | A,B | 0.1 |
| Current sensing | B | 0.1 |
| Self heating of the coil | A,B | (origin 56 →) 7 |
| Force weighing | A,B | (origin 20 →) 1 |
| Mass measurement | B | (origin 5→) 0.25 |
| Length measurement | A,B | (origin 49 →) 0.5 |
| Gravity | A,B | 0.01 |
| Air buoyancy | A,B | 1 |
| Combined uncertainty | A,B | (origin 77→) 7.2 |

### H. The Future Plan of NIM-1

The NIM-1 is being improved further. At the same time, it is also used to do some model test for the design and optimization of the new joule balance NIM-2.

### III. THE DESIGN AND CONSTRUCTION OF THE NEW JOULE BALANCE NIM-2

To make contribution for the final redefinition of kilogram, the uncertainty has to be less than 0.05ppm [9]. But for NIM-1, there are three aspects that limit its uncertainty to $10^{-6}$ level. The first is its big size and it cannot be put into a vacuum chamber. The length measurement will be limited to $10^{-6}$ level in air environment. The 2$^{nd}$ limitation lays on the resolution and short time repeatability of the balance. The 3$^{rd}$ one is the self heating of the coils with currents.

Thus since early 2013, the design of a brand new joule balance has been under consideration. The criteria lay on the following, 1) The new apparatus should be a compact and vacuum compatible system. Thus the size has to be decreased a lot compared with NIM-1. 2) A better weighing device is needed to replace the 2-meter long balance in NIM-1. The resolution and repeatability should be at micrograms level for a 1kg mass. 3) The self-heating of the coils should be decreased a lot. Otherwise, it cannot be used in a vacuum chamber. With these considerations, the design of the NIM-2 is in progress and details are presented as following.

*A. Compact Size and Vacuum Compatible Consideration*

A commercial mass comparator has been ordered, thus the size of the whole apparatus can be decreased a lot.

Without oil damping device mentioned above, the laser based position locking system to keep the suspended coil stable cannot work well. However, the oil damping device will bring big trouble in vacuum environment. A 3-dimensional magnetic damping device is proposed by authors as shown in Fig.10 (a). It is a magnetic multi-pole combined with a U-type iron yoke and 8 pieces of permanent magnet as shown in Fig. 10 (b). The magnetic lines of the damping device in z direction and y direction are shown in Fig. 10 (c) and Fig. 10 (d) separately. Experiments show that it works well. In theory, its leakage magnetic field attenuates quickly in the surrounding area.

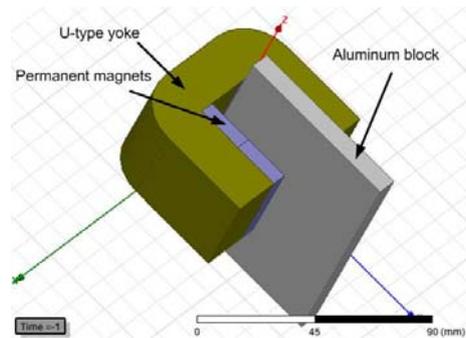

(a) The structure of the 3-dimensional magnetic damping device

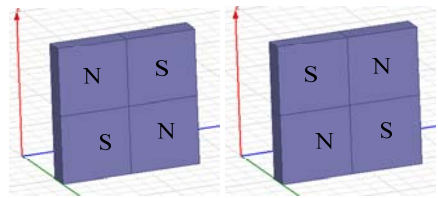

(b) The magnetic pieces in both sides of the yoke (left and right)

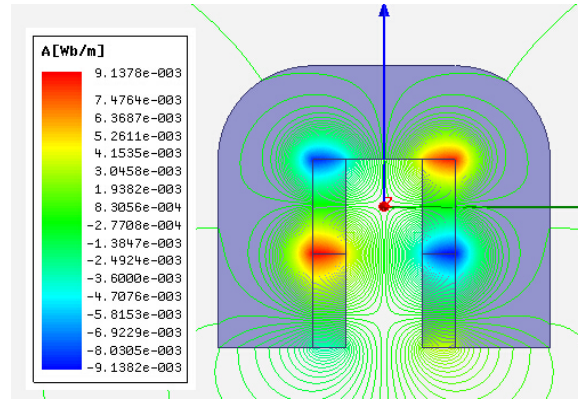

(c) The magnetic line of the damping device in z direction

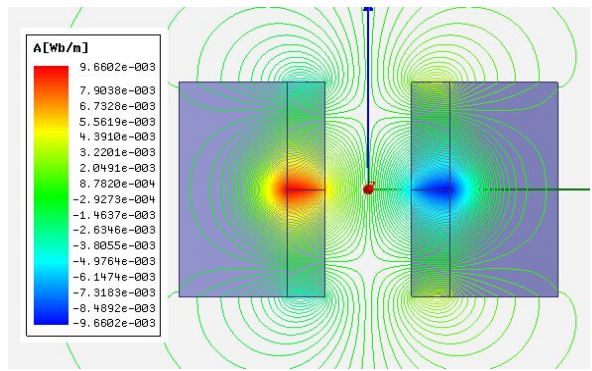

(d) The magnetic line of the damping device in y direction
Fig.10. The 3-dimensional magnetic damping device

*B. Mass Comparator*

A commercial mass comparator with 5kg measurement capacity has been ordered to replace the long beam balance. Its resolution is 1μg and repeatability is 15 μg. Thus, in the weighing phase, the uncertainty can be decreased to 0.02ppm for the 5N electrical force. The reason for choosing a 5kg capacity is that the new suspended system including suspended coil is almost 4.5kg. In the weighing phase, at the beginning, a 1kg mass is loaded on. At the same time, a 5N electrical force in up direction is generated to keep the total mass of the suspended system close to 5kg. In the next step, when the 1kg mass is removed, a 5N electrical force in down direction is generated to let the mass comparator almost feel no change. In the second phase, the mass comparator is reset to 5kg by loading a 0.5kg mass on the suspension. This will keep the mass comparator under the same load in both phases.

*C. Electromagnet*

To decrease the self-heating of the exciting coils further, an electromagnet using ferromagnetic material has been designed as shown in Fig. 11 [15]. The idea is from the consideration that if the ferromagnetic material is used to short the magnetic path between the exciting and movable coils, the efficiency of the magnetic field can be increased a lot. At the same time, this design is a close circuit, thus the shielding is very good. Fig.11 shows the structure of the new electromagnet.

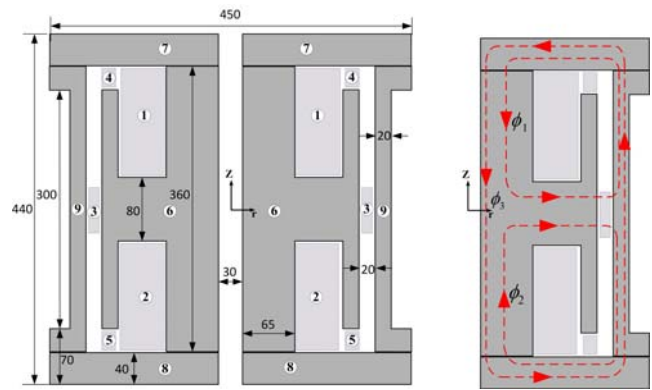

Fig.11 the structure of the new electromagnet

The materials in the Fig.11 are as following,
   -1 and 2: exciting coils;
   -3: the suspended coil;
   -4 and 5: the compensation coils for the suspended oil;
   -6, 7, 8 and 9: the soft iron (DT4C type).

Two fixed coils 1 and 2 with the same ampere-turns are connected in opposite to be the exciting coil. Number 3 is the suspended coil in the air gap; coils 4 and 5 are compensation coils. The number of turns of every compensation coil is equal to the half of the turns of the suspended coil 3. The role of compensation coils is shown in Fig.12 and is explained later.

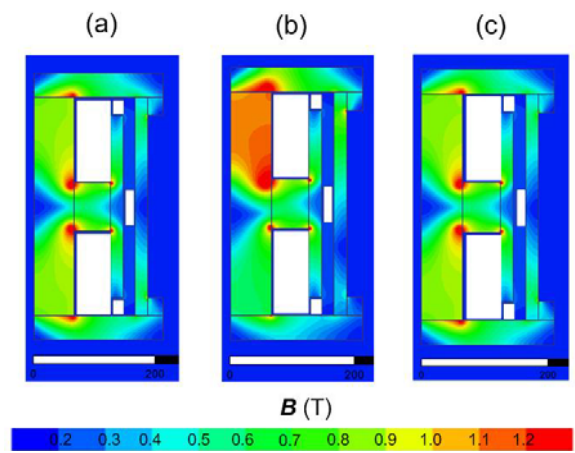

Fig.12. Demonstration of the function for compensation coils. (a) The magnetic field offered by currents in exciting coils only. (b) The magnetic field given by currents in exciting and suspended coils, but current in compensation coils is 0. (c) The magnetic field given by currents in exciting and suspended coils, also compensation coils. Note that in three cases of the simulation a 0.05mm installation gap between the cover and the main yoke (inner vertical core/outside cylinder) is applied.

With current in exciting coils 1 and 2 only, the simulation of the magnetic flux density in the magnet cores is shown in Fig.12(a) and no saturation is observed. However, when the suspended coil is with current, saturation appears in many parts of the magnet cores as shown in Fig.12(b). The reason is from more magnetic flux generated by the suspended coil 3 along the closed magnetic circuit $\varphi_3$ where no air gap exists and the magnetic reluctance is small. If the magnetic property of the materials is absolutely linear, the saturation of core doesn't disturb the system operation. However, the ferromagnetic substance is actually nonlinear, and it will lead to different magnetic states of magnet in different operating modes, thus systematic errors will increase. Therefore, the compensating coil 4 and 5 with the same number of turns as the coil 3, are electrified to generate magnetic potential in the opposite directions to that generated by the coil 3. In this case, no magnetic potential appears along the magnetic circuit $\varphi_3$. The simulation of this situation is shown in Fig.12(c), and it is obvious that the magnetic states are more perfect.

For the new electromagnet, the current for the exciting coil is 0.5A and the magnetic field in the gap (about 0.08 Tesla), in which the suspended coil is located, is much stronger than the case in NIM-1. To generate an electrical force of 5N, the parameters of the electromagnet are shown in Table III.

TABLE III
THE PARAMETERS OF THE NEW ELECTROMAGNET

| Item | Unit | Value |
|---|---|---|
| **Exciting coils** | | |
| Resistance | ohm | 37 |
| Current | A | 0.5 |
| Power | W | 9.25 |
| **Compensation coils** | | |
| Resistance | ohm | 3.4k |
| Current | mA | 7 |
| Power | W | 0.17 |
| **Suspended coils** | | |
| Resistance | ohm | 4.1k |
| Current | mA | 7 |
| Power | W | 0.2 |

Here the current for the suspended coil is about 7mA, and the electrical force is close to 5N, thus a 1kg mass can be used.

The temperature of the new magnet with a current of 0.5A is measured as shown in Fig.13. The temperature increase of the air space for suspended coil is about 1 ℃. To decrease the temperature further, the heat tube is in consideration.Besides, the range of moving distance is enlarged from 20mm to 100mm. Thus the uncertainty from the length measurement could also be reduced a lot.

For the coils used in NIM-1, the mutual inductance is measured with the standard square-wave compensation method proposed by authors [8]. For the new electromagnet with ferromagnetic core to be used in NIM-2, the mutual inductance is very big. It is very difficult to measure it with high precision. Besides, the current in the suspended coil has to be changed and the non-linear characteristic and hysteresis will bring a big problem. New approach named the magnetic flux linkage measurement is proposed and details are presented in [15].

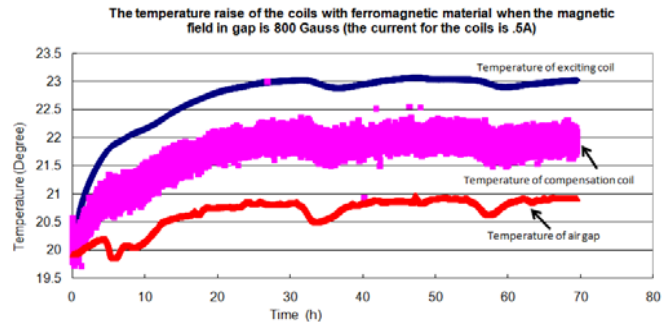

Fig. 13. Temperature of the new magnet with a current of 0.5 A

The advantages of the new magnet are as following, a) The new magnet is a close circuit. The outer ring acts as the magnetic shielding. Thus the disturbance from environment can be decreased a lot. b) The electromagnetic force is increased to 500g, which is 2.5 times of that in NIM-1. c) The self-heating is decreased greatly. The power of the exciting coil is less than 10W. d) The homogeneity is determined by the geometry of the air space and easy to be realized. The height of the homogeneity area is enlarged to 100mm, which is 5 time longer than before.

Of course, there are also some disadvantages for the new magnet with soft iron. The weight of the new magnet is almost 500kg and it is difficult to move it up and down. However, if the suspended coil is moved with the mass comparator, the repeatability of the mass comparator will be degraded a lot. Finally, in NIM-2, the mass comparator with suspension system is kept at a stable stage and the electromagnet will be moved to different position along a straight track in the gravity direction. A translation stage moving the electromagnet has been design and is being fabricated by a team from Harbin Institute of Technology and will be ready by the end of 2014.

Besides, to prevent the vibration of surroundings, an active control vibration isolation base is designed and in fabrication. To keep the suspended coil from swinging and vibration, the laser system for position measurement and locking of the suspended coil, which has been proposed for the NIM-1, together with above mentioned 3-dimentional magnet damping device, will be used in the NIM-2.

The overview of the NIM-2 is shown in Fig.14.

At present, the mass comparator, electromagnet and the laser system for position measurement and locking of the suspended coil are ready. The other 3 parts, including vacuum chamber, the active control vibration isolation base and the translation stage for the electromagnet are still in research and fabrication. They will be ready by the end of 2014 or a little late. The whole system is expected to be mounted by the end of 2015 and be ready for testing.

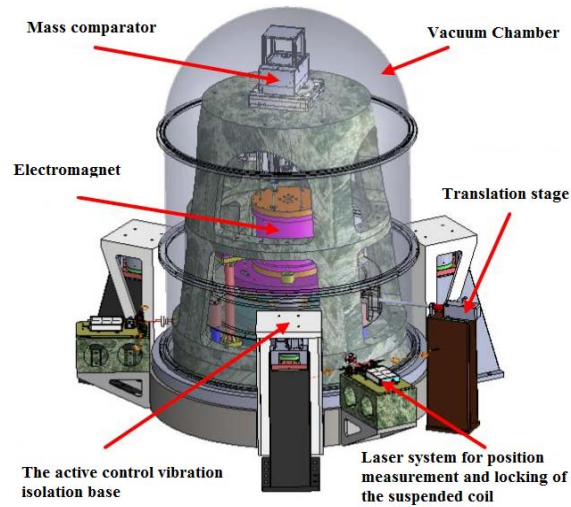

Fig.14. The overview of the new joule balance NIM-2

## IV. CONCLUSIONS

The joule balance method to measure the Planck constant and redefine the Kilogram ultimately has been going on at the National Institute of Metrology of China (NIM) since 2007. The first prototype has been built to check the feasibility of the principle. In 2012, the relative uncertainty of the Planck constant measurement result is at $7.7\times10^{-5}$. Self-heating of the coils, swing of the coil, are the main uncertainty contributions. Since 2012, some improvements have been done to decrease these uncertainties. The relative uncertainty of the joule balance is decreased to $7.2\times10^{-6}$ at present. The Planck constant measured with joule balance is $h=6.6261041(470)\times10^{-34}$Js. The relative difference between it and CODATA2010 recommendation value is $5\times10^{-6}$. Further improvements are still being carried out on the NIM-1 apparatus. At the same time, the design and construction of a brand new and compact joule balance NIM-2 are also in progress. The approaches to decrease the uncertainty contributions are described here.

## V. ACKNOWLEDGMENT

Authors would like to give thanks to Dr. Jiubin Tan and his team for their excellent contribution for the design and fabrication of the translation stage, the active control vibration isolation base and the laser system for position measurement and locking of the suspended coil, as well as the machining and the mounting of the new electromagnet. Thanks also would like to be given to Mr. Hongzhi Gong and Mr. Jinduo Han for their perfect work in the machine shop for the coil winding work.